\documentclass[conference]{IEEEtran}
\IEEEoverridecommandlockouts
\usepackage{cite}
\usepackage{amsmath,amssymb,amsfonts}
\usepackage{graphicx}
\usepackage{textcomp}
\usepackage{xcolor}
\usepackage{booktabs}
\usepackage{multirow}
\usepackage{tabularx}
\usepackage{hyperref}

\usepackage{array}
\newcolumntype{P}[1]{>{\centering\arraybackslash}p{#1}}
\newcolumntype{M}[1]{>{\centering\arraybackslash}m{#1}}

\hypersetup{
	colorlinks=true,
	linkcolor=blue,
	citecolor=black,
	filecolor=black,
	urlcolor=blue,
}
\graphicspath{ {images/} }
\def\BibTeX{{\rm B\kern-.05em{\sc i\kern-.025em b}\kern-.08em
    T\kern-.1667em\lower.7ex\hbox{E}\kern-.125emX}}
\begin{document}

\title{Designing a Sequential Recommendation System for Heterogeneous Interactions Using Transformers  \\
{\footnotesize \textsuperscript{}}
}
\author{\IEEEauthorblockN{Mehdi Soleiman Nejad, Meysam Varasteh, Hadi Moradi and Mohammad Amin Sadeghi \\}
\IEEEauthorblockA{School of Electrical and Computer Engineering\\
University of Tehran \\
Tehran, Iran \\ \\
m.soleimannejad@ut.ac.ir \\
meysamvaraste@ut.ac.ir \\
moradih@ut.ac.ir \\
asadeghi@ut.ac.ir \\}}

\maketitle

\begin{abstract}
While many production-ready and robust algorithms are available for the task of recommendation systems, many of these systems do not take the order of user's consumption into account. The order of consumption can be very useful and matters in many scenarios. One such scenario is an educational content recommendation, where users generally follow a progressive path towards more advanced courses. Researchers have used RNNs to build sequential recommendation systems and other models that deal with sequences. Sequential Recommendation systems try to predict the next event for the user by reading their history. With the massive success of Transformers in Natural Language Processing and their usage of Attention Mechanism to better deal with sequences, there have been attempts to use this family of models as a base for a new generation of sequential recommendation systems. In this work, by converting each user's interactions with items into a series of events and basing our architecture on Transformers, we try to enable the use of such a model that takes different types of events into account. Furthermore, by recognizing that some events have to occur before some other types of events take place, we try to modify the architecture to reflect this dependency relationship and enhance the model's performance.\end{abstract}

\begin{IEEEkeywords}
Recommendation Systems, Transformers, Deep Learning
\end{IEEEkeywords}

\section{Introduction}
With the massive amount of information available to us in the snap of a finger, having a system that finds useful and relatable content for us is of paramount importance. Recommender systems have been trying to predict the user's interests by using the information available to us, mostly in the form of likes and dislikes by the user \cite{10.1145/245108.245121}. While many of these systems don't account for the change of taste users have over the course of time, there has been a push for modeling dynamic sequential behavior in newer recommender systems thanks to the progress made in Deep Learning \cite{cheng2016wide} \cite{10.1145/2959100.2959190} \cite{10.1145/3159652.3159656} \cite{10.1145/3159652.3159668} and other fields \cite{ijcai2019-883}. There are many instances where the order of interaction with items should matter to make a good recommendation. One of the areas the users might benefit greatly from this sequential modeling is educational content. Due to the progressive nature of many courses where certain topics are covered by them with different levels of complexity, the order of the consumption by the user matters. For example,  while "Introduction to Programming" and "Advanced Programming" are highly correlated courses and generally draw the same audience of users, it is unwise to recommend "Advanced Programming" to a user that has not passed "Introduction to Programming". The user would be confused when taking the course, because many topics from the other course are needed to fully understand the advanced material, and it is likely the user will leave the course before finishing it, thus lowering their engagement with the platform. Similarly, we should not recommend "Introduction to Programming" to a user that has passed "Advanced Programming". It is unlikely that there are any new material in the course for this user, as they have passed the more advanced course. 

To model such behavior, various methods have been proposed \cite{kang2018selfattentive} \cite{ijcai2019-883}, focusing on predicting the user's next interaction given their interaction history. Many of these methods integrated Recurrent Neural Networks (RNNs) such as \cite{10.1145/3109859.3109877} \cite{hidasi2016sessionbased}, \cite{10.1145/3269206.3271761}. These methods try to capture the essence of a user's experience into a hidden state vector and predict the next most likely item with said vector. While these methods have been effective in tackling such modeling, with the rise of the Transformers family \cite{vaswani2017attention} and specially BERT \cite{devlin2019bert} breaking several records in Natural Language Processing (NLP) tasks, researchers have focused on using this family of deep learning models to accommodate tasks that are not traditionally aligned with NLP. A good example would be a recommender system designed with the mechanisms introduced by the Transformers family \cite{sun2019bert4rec} \cite{chen2019behavior} \cite{10.1145/3340531.3411954} \cite{10.1145/3383313.3412258}. Furthermore, tech heavyweights Alibaba have also adopted Transformers for recommendation recently \cite{chen2019behavior}. The main idea behind using Transformers for the task of recommendation is to create sequences from the interactions each user has with items, and then use the same mechanisms to predict the next item that is most likely to appear in the sequence. We can think of these sequences as sentences and each item in the sequence as a word. 

While these models are sophisticated and out-perform their recurrent peers, they are often designed to work with homogeneous events. For example, \cite{sun2019bert4rec} completely disregards the negative reviews and only uses positive reviews to form a profile for each user and uses this profile to predict the next item the user will like. This limitation makes the model an unideal solution for scenarios where predicting other events can also be helpful. We want to have a model that can consider different types of interactions as well. These events can include typical user interactions such as leaving a positive or negative review, registering for a course, or passing it. It is also possible for these events to be dependent on each other. For example, users can't review a course they have not registered for in many learning platforms, so the events can also form a sequential order. This means we will have to train a model for each type of event if we want to accommodate different events into our recommendation. Another direction for improvement that we want to take is to go beyond simple hyperparameter tuning and slightly modify the architecture to accommodate specific properties we introduce to our input via preprocessing. 

In this work, We first use a simple technique to create heterogeneous sequences of events that follow certain grammar-like rules for each user from the user-item interaction matrix. We then modify BERT by altering the architecture to accommodate the grammar of our synthetic language that represents our data. This means that the model will have the structural limitations of our synthetic language as some implicit hyperparameter, and there is no need to infer them from the dataset and learn them during the training. This allows the model to use its learning capacity to infer other useful information from the dataset. Similar to previous models, we also use the Masked Language (Cloze Test) task to train and evaluate our model. Ultimately, we measure how much our modifications improve the results compared to the original architecture and then compare our model to other sequential models. Due to the method of our transformation of data into sequences, our proposed model will also be able to predict the type of events for certain items in addition to the item related to the event.

The rest of this paper is organized into four sections. In section \ref{section-2}, we briefly review related works. In section \ref{section-3}, we present our model with details. In section \ref{section-4}, we evaluate our proposed model on an online learning platform's dataset and compare it with other models. Ultimately, in section \ref{section-5} we conclude our work with pointers for future directions.

\section{Related Work}\label{section-2}
In this section, we discuss some previous work done in recommender systems briefly.

Recommendation systems have been a field of interest for researchers since the '90s. Many researchers have opted to study Collaborative Filtering (CF) \cite{Schafer2007}, a family of methods that tries to learn each user's preferences by finding similarities between users and items using the User-Item Matrix. In this matrix, each row represents a user and each column an item. Since it's unlikely that users interact with the majority of items, this matrix is sparse. The most popular CF method is Matrix Factorization \cite{NIPS2007_d7322ed7} \cite{10.5555/1005332.1044709}, in which the algorithm tries to break the matrix into two separate matrices, one for representing a hidden state for each user and one for items. Due to the rise of Deep Learning, there has been immense progress in creating better recommender systems \cite{10.1145/3038912.3052569} \cite{10.1145/3097983.3098077} \cite{Xiao_Liang_Shen_Meng_2019} \cite{varasteh2021improved}. The progress of deep learning meant that it was now possible to create more complex recommender systems that can use different sources of data, such as text \cite{10.1145/2959100.2959165}, images \cite{DBLP:conf/icdm/KangFWM17} or sound \cite{NIPS2013_b3ba8f1b} \cite{10.3389/fams.2019.00044} to find the preferences each user has. In addition to the advanced in using deep learning for a recommendation, some methods also use hybrid techniques to utilize different models\cite{8734976} better.

Generally speaking, CF methods have no sequential nature. They do not consider the order of interaction when predicting an item's score, and they treat every given score as if they are given at the same time. As argued previously, the order of user-item interaction can be good information to have on certain use cases. Markov Chains (MCs) \cite{10.1145/1772690.1772773}, Factorization Machines \cite{9018047} \cite{10.1145/3240323.3240356} and Recurrent Neural Networks (RNNs) \cite{10.1145/3269206.3271761} are good choices for dealing with sequential data and there has been successful efforts to use such architectures to model a sequential recommender systems. Researchers have used MCs to create a graph to model the user's sequential behavior and combine this with other means to predict the next item.
Meanwhile, the main idea in RNN-based recommender systems is to convert a user's history into a hidden state vector and use that vector to predict the next item. Some architectures use Convolutional Neural Networks (CNNs) to model the sequence \cite{10.1145/3159652.3159656} \cite{10.1145/3109859.3109900} where the sequence is thought of as some image that we have to extract features. Apart from trying new architectures for better sequential recommendations, there has also been researching for providing model-agnostic frameworks for sequential recommendation \cite{10.1145/3404835.3462908}. 
After the introduction of Attention mechanism and Transformer family \cite{vaswani2017attention} and later on BERT \cite{devlin2019bert} by Google, the research team behind BERT managed to break several records in NLP. This major breakthrough in NLP inspired other researchers to use the new techniques in other fields that might benefit from modeling sequences. 

While BERT itself is a pre-trained model for language processing that can be fine-tuned to certain datasets or tasks, the architecture can be used in different scenarios like sequential recommendation systems. SASRec \cite{kang2018selfattentive}, BST \cite{chen2019behavior}, BERT4Rec \cite{sun2019bert4rec} and S\textsuperscript{3}-Rec \cite{10.1145/3340531.3411954} are some of the better known models to utilize Attention mechanism and ultimately Transformers for sequential recommendation.
\section{Architecture}\label{section-3}
\begin{figure*}[h]
	\centering
	\includegraphics[width=0.8\textwidth]{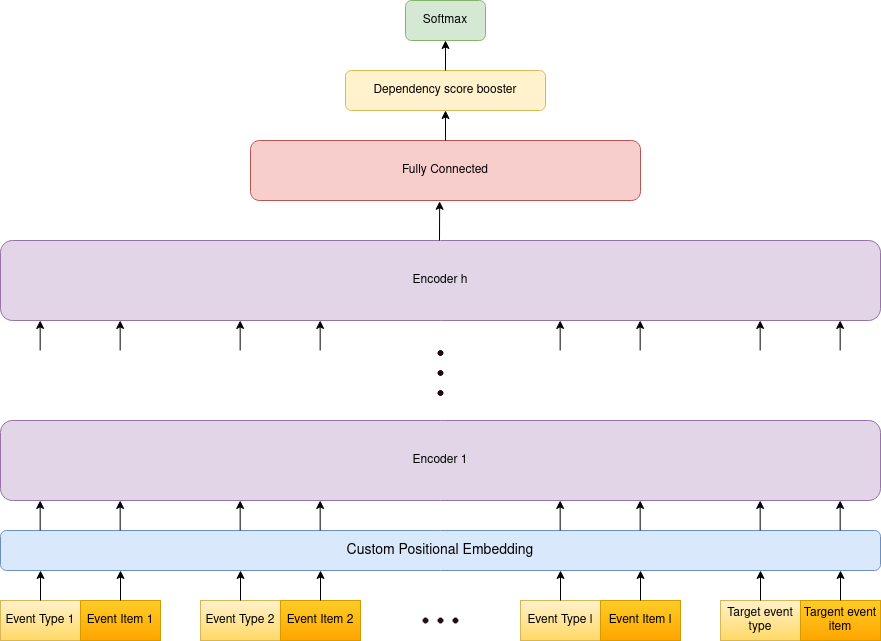}
	\caption{The overview of the proposed model}
	\label{fig:architecture}
\end{figure*}
In this section we define our proposed architecture to design a sequential recommender system that deals with heterogeneous item history.
\subsection{Problem Statement}
In our task, suppose \(U = \{ {u_{1}, u_{2}, ..., u_{N}}\} \) is the set of the users, \(I = \{i_{1}, i_{2}, ..., i_{M}\}\) is the set of the items and also \(E = \{e_{1}, e_{2}, ..., e_{K}\}\) is the set of possible events and actions that each user might perform on an item. We acknowledge that certain events might only happen after another certain event has taken place. For example, it is common practice to only let users who have bought the product to leave a review on it. We define \(D(u, \hat u)\) as a function that defines this relation between our events. We then define \(S(u) =\{(e,i)_{1}, (e,i)_{2}, ..., (e,i)_{l} \}\) where \(u \in U\), \(e \in E\) and \(i \in I\) as the interaction history of user \(u\). These interactions are sorted by their time of occurrence. Given the interaction history \(S(u)\) and either \(e_{l+1}\) or \(i_{l+1}\), our recommender system should predict the missing half of event \(l+1\). The proposed model is shown in fig. \ref{fig:architecture}.

\subsection{Learning}
\subsubsection{Train/Test Data Split}
The data split we perform a bit different from the traditional methods. For each user, we save the last item in their sequence \(S(u)\) as the test item and use the rest during training. This means that any user with at least two different interactions in their history is used in the training process. 
\subsubsection{Training}
\begin{figure}[h]
	\centering
	\includegraphics[width=0.35\textwidth]{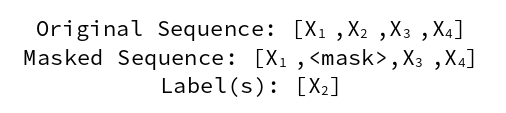}
	\caption{Masked Language example}
	\label{fig:cloze-test}
\end{figure}

We use the Masked Language task to train our model. In this task, some of the tokens in the sequence are randomly masked, and the model predicts the missing token. An example of this is shown in fig. \ref{fig:cloze-test}. In this example, the model should fill in the space denoted by \(<mask>\) with \(X_{2}\). This is done by adding an output layer (also known as a head layer) that maps the transformer's output to a distribution over all possible tokens using a fully connected neural network and applying Softmax. We do many iterations and randomly mask tokens again each time, so all tokens get a fair chance to be masked and impact the tunable parameters.
\subsubsection{Testing}
We can do two types of testing on our model. In the first type, we give the user history and the event type \(e_{l+1}\) and mask the item ID. The model will then predict the appropriate item ID for the given event type and history. We can output a list of candidates to the model that has given a high score. In the second type, we mask the item ID \(i_{l+1}\) instead, and the model should predict the appropriate event type for the item.
\subsection{Data Preprocessing}
\begin{figure}[h]
	\centering
	\includegraphics[width=0.35\textwidth]{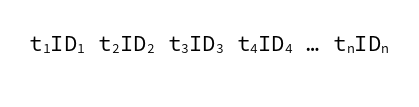}
	\caption{The structure for each user sequence}
	\label{fig:data-structure}
\end{figure}

While most user-item interaction datasets are usually tabular, to feed our data to BERT, we need to convert the data into sequences for each user. We present a simple technique to do this transformation. For each user, we gather all interactions and sort them by the time of their occurrence. We treat each interaction as an event. We decide on unique symbols for each type of event, such as like or dislike. We then concatenate the event's symbol with the item's ID to represent this interaction. This makes each of our sequences follow the structure shown in fig. \ref{fig:data-structure} where \(t_{i}\) is the type and \(ID_{i}\) is the associated item ID for event \(i\). We then create a tokenizer that separates \(t_{i}\) from \(ID_{i}\) and converts the user-item history to a sequence of integers, ready to be given as input to the embedding layer. This means that we have effectively transformed \(S(u)\) into \( \{ e_{1}, i_{1}, e_{2}, i_{2}, ..., e_{l}, i_{l} \}\), where each \(t_{n}\) is representing \(e_{n}\) and each \(ID_{n}\) is representing \(i_{n}\).

\subsection{Embedding Layer}
Our first component is the embedding layer, which projects each part of our sequence to fixed-size vectors. These vectors represent their corresponding tokens and are fed into the next layer. In \cite{vaswani2017attention}, the writers have suggested Positional Embedding as their choice for this layer. That means for a given item \(i_{k}\), we will use the following representation:
\begin{equation}
  h_{i} = v_{i} + p_{i}
\end{equation}
where \(v_i\) is the fixed-dimensional embedding of \(i_{k}\) and \(p_i\) is the positional embedding of \(i_{k}\) and of same dimensionality of \(v_i\). 

This choice is to reflect the order of the items in the projections. The writers notice that there is minimal difference between using a fixed sinusoid implementation or a tunable layer. We opt to use a tunable layer for convenience, but we leave two elements from the vector out of the training process and set them manually. These two elements each emphasize a different structural limit in of \(S(u)\).
As stated in the previous section, if we break \(S(u)\) into \( \{ e_{1}, i_{1}, e_{2}, i_{2}, ..., e_{l}, i_{l} \}\), we quickly notice that the type of tokens in all the odd positions are different from the type of tokens in the even positions. That is, we only have event types in odd positions and item IDs in even positions. We use one of the elements to reflect this fact and label each token as either an event or an item. This will help the model realize the difference quickly.
\begin{equation}
P_{type}(i)= 
\begin{cases}
    0, & \text{if \(i = 2k\)}\\
    1, & \text{if \(i = 2k + 1\)}
\end{cases}
\end{equation}
After breaking the tuples in \(S(u)\), we have to find a new way to show the relationship between each item ID and event type. Using another element and setting it to the same value for the related event type and item ID token, we label them together. The model will have a strong clue to figure out the relationship between each event type and item ID.

\begin{equation}
P_{relevance}(i)= 
\begin{cases}
    sin(\omega_{k}.i), & \text{if \(\left \lfloor{\frac{i}{2}}\right \rfloor = 2k\)}\\
    cos(\omega_{k}.i), & \text{if \(\left \lfloor{\frac{i}{2}}\right \rfloor = 2k + 1\)}
\end{cases}
\end{equation}

This slight change helps the model to be more accurate by effectively labeling each token of the sequence.
\subsection{Transformer Layer}
\begin{figure}[h]
	\centering
	\includegraphics[width=0.35\textwidth]{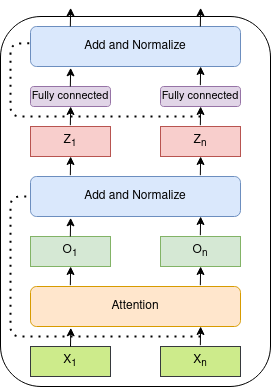}
	\caption{The Encoder layer}
	\label{fig:encoder}
\end{figure}
The second component is the Transformer layer, which learns a deeper context for each token by analyzing its relationship with other tokens in the sequence. This happens by stacking Encoder modules on top of each other. The structure of each Encoder is shown in fig. \ref{fig:encoder}. 

The input to this component is \(h_i\) produced by the embedding layer. The input for each layer \(l\) is denoted by \(h^{l}_{i}\). By stacking these hidden vectors, we create the matrix \(H^l\) for denoting the input for each layer. Each Encoder consists of two main parts, which, as shown in fig \ref{fig: Encoder} we use with residual connections followed by a normalization layer. 1) Multi-Headed Attention and 2) Point-Wise Feed Forward.
\subsubsection{Multi-Headed Attention}
The attention mechanism is at the heart of the Transformer architecture and is regularly used to capture relationships and dependencies in sequences, regardless of their relative distance to each other. 
Attention is calculated as follows:
\begin{equation}
  Att(Q, K, V) = softmax(\frac{QK^T}{\sqrt{d}})V
\end{equation}

Where \(Q\) is the query matrix, \(K\) is the key matrix, \(V\) is the value matrix, and \(d\) is the dimension of each key vector. These matrices are linear projections from the original \(h_i\) produced by the previous layer.  As discussed in \cite{vaswani2017attention}, it is beneficial to use different representation spaces together instead of just using one attention module. We'll adopt the Multi-Headed Attention mechanism:

\begin{equation}
  MHA(H) = concat(head_1(H), ..., head_n(H))W^{MH}
\end{equation}
\begin{equation}
  head_i(H) = Att(HW^{Q}_{i}, HW^{K}_{i}, HW^{V}_{i})
\end{equation}
where \(W^{MH}\), \(W^{Q}_{i}\) , \(W^{K}_{i}\) and \(W^{V}_{i}\)  are tunable parameters.

\subsubsection{Point-Wise Feed Forward}

Since all the attention operations are linear projections, we need to introduce non-linearity to enhance the model. We do this by using adding a point-wise feed-forward layer on top of our multi-head attention layer.

\begin{equation}
  PWFF = concat(FF(h_{1}^{l}), ..., FF(h_{n}^{l}))
\end{equation}
\begin{equation}
  FF(h) = GeLU(hW^{(1)} + b^{(1)})W^{(2)} + b^{(2)}
\end{equation}
where \(W^{(1)}\), \(W^{(2)}\) , \(b^{(1)}\) and \(b^{(2)}\)  are tunable parameters shared across different points. Following BERT \cite{devlin2019bert}, we opt to use GeLU as our activation function instead of ReLU.

\subsubsection{Stacking Encoders}
By stacking encoders on top of each other and feeding the output of the previous Encoder to the next layer, we introduce further complexity to our model. We use dropout at the output of each layer to avoid overfitting and introduce generalization to our model. The output of the last Encoder is used as input for our final component, the output layer.

\subsection{Output layer}
The last component of our architecture is the output layer. This layer is a fully connected neural network that maps the context vector for the masked items over target items or events. Using a two-layer fully connected neural network, we give each token a score for every masked item.
As noted before, since some events can be dependent on other events, we utilize this knowledge to boost our model's accuracy by limiting the number of possible candidates. If the masked token is an item ID (i.e., odd-positioned), we check if the event type has any dependency. If there is a dependency on another event type, we increase the score for all the item IDs with said dependency by 1. This allows our model to differentiate between irrelevant and valuable items and enhance our model. We then apply the Softmax function over the scores to distribute the target item IDs and event types.
Ultimately, the token with the highest probability will be chosen as the prediction for the masked item. 

\section{Experiment}\label{section-4}
This section talks about our dataset, the metrics used for experimenting, and our configuration for the model. We then measure the impact of our proposed changes in helping the Transformer model predict the next item. Ultimately, we compare our proposed models against other models.
\subsection{Dataset}
One of the best examples of sequential behavior is online learning platforms and digital courses due to the sequential nature of learning topics. The dataset used for this paper is from an online learning platform. In this dataset, users register for courses and either like the course or dislike it. We use \(r\), \(+\) and \(-\) as tokens to represent these events respectively. We know that liking and disliking a course is also dependent on registering for the course first. After preprocessing the dataset, there are about 47000 users available to us. The total number of courses is 630.
\subsection{Evaluation Metrics}
As previously stated, there are two types of tests we can perform using our model. In the first type where the model predicts the item ID, we use HR@1 (Hit Ratio), HR@5, HR@10, NDCG@5 (Normalized Discounted Cumulative Gain) NDCG@10. For our specific task, the sequential recommendation, NDCG@1, and HR@1 are equal, so we only report HR@1. We also use these metrics to compare our model with other models.
In the second type, where the model has to predict the event type, we use precision, recall, and F1-score and treat the test as a classification since there are three types of events. We'll create a 4th class that does not have any correct labels. We use this 4th class to label anything the model produces that is not associated with the types.
\subsection{Configuration}
\begin{table*}[h]
\def\arraystretch{2.5}
\caption{Performance comparison of different models}
\begin{center}
\begin{tabular}{ | M{5em} | M{7em} | M{7em} | M{7em} | M{7em} | M{7em} | M{7em} |} 
\hline
Metric & POP & End-To-End BERT & SASRec & BERT4Rec & Proposed Model & \textbf{Improvement} \\
\hline
HR@1 & 0.029 & 0.086 & 0.111 & \underline{0.135} & 0.150 & 11.1\% \\
\hline
HR@5 & 0.099 & 0.320 & \underline{0.464} & 0.303 & 0.512 & 10.3\% \\
\hline
HR@10 & 0.152 & 0.489 & \underline{0.564} & 0.521 & 0.624 & 5.9\% \\
\hline
NDCG@5 & 0.063 & 0.208 & \underline{0.311} & 0.220 & 0.344 & 10.6\% \\
\hline
NDCG@10 & 0.080 & 0.251 & \underline{0.356} & 0.255 & 0.378 & 6.1\% \\
\hline
\end{tabular}
\end{center}
\label{tab:comparison}
\end{table*}
For this experiment, we'll use eight heads for the attention module and stack two layers of encoders on top of each other. Our context vectors' length will be 96. The output layer is set to be 128. We mask \(p=0.4\) of the tokens randomly and rerun the training algorithm for 50 epochs with batch sizes of 256. Other models will also feature similar complexity for a fair comparison. We use the excellent implementations provided by HuggingFace using PyTorch as our base and apply our modifications to their code. Ultimately, we train our model on a machine using a single Nvidia GeForce GTX 1070 GPU fully from scratch, without any pretraining.
\subsection{Performance Comparison}
We compare our models to the following baselines:
\begin{itemize}
  \item \textbf{POP}: This naive method ranks items according to their popularity and always recommends them to each and every user. 
  \item \textbf{End-To-End BERT} \cite{devlin2019bert}: We use our preprocessed data to train BERT from scratch without any pretraining and only using our custom tokenizer. This means that none of our proposed changes are applied, and we only use our heterogeneous data.

  \item \textbf{SASRec} \cite{kang2018selfattentive}: This method uses a unidirectional Transformer model to predict the next interaction with items.
  \item \textbf{BERT4Rec} \cite{sun2019bert4rec}: This method uses a bidirectional Transformer model, BERT, to predict the next interaction with items. 
\end{itemize}

With the sole exception of POP, the other methods use Transformers in their architecture. Table \ref{tab:comparison} shows the scores across different metrics for these models, our proposed model, and the improvement our model has achieved over the best performing alternative. In each row, the best score for alternative models is underlined.  As is evident in table \ref{tab:comparison}, POP is predictably the worst performer among the models due to a lack of personalization, while the best performance among alternative models belongs to SASRec, only losing to BERT4Rec when measured using HR@1. 
Compared to the best performing model in each measurement, our model sees an improvement close to 10\% and 6\% when the output list size is 5 and 10, respectively. We see the most significant improvement for HR@1 with 11.1\% and the slightest improvement for HR@10 with 5.9\%. 
\subsection{Effect of custom positional encoding in event type classification}
\begin{table}[h]
\def\arraystretch{2.5}
\caption{Performance comparison of different models}
\begin{center}
\begin{tabular}{ | M{5em} | M{7em} | M{7em} | M{6em} |} 
\hline
 & W/ custom encoding &  W/o custom encoding & \textbf{Improvement} \\
 \hline
Precision & 0.685 & 0.678 & 1.0\% \\
\hline
Recall & 0.577 & 0.555 & 3.9\% \\
\hline
F-1 Score & 0.585 & 0.565 & 3.5\% \\
\hline
\end{tabular}
\end{center}
\label{tab:classification}
\end{table}
We now study the effect of our proposed positional encoding on correctly classifying the event type for each event. To study this, we train our model once with our proposed positional encoding and once with completely tunable positional encoding from scratch, without any pretraining and any changes over other parameters. We then use the trained models to predict the event type of the next interaction. Table \ref{tab:classification} shows the results of this experiment. As evident, precision does not improve much but recall, and F-1 scores are improved 3.9\% and 3.5\% respectively. 
\section{Conclusion and Future Work}\label{section-5}
This paper presented a method for using the Transformer family of models for sequential recommendation systems. Instead of focusing on only one type of feedback from the user, such as likes, by simply adding a token to differentiate between events, we provide a framework that allows other types of events to be taken into account. We then try to exploit the fact that many events can only happen if a particular event has happened before them in real life, such as a user having the ability to positively or negatively review an item only after ordering that specific item. By modifying the structure of the model to factor in this rule and event types, we introduce a model that outperforms other models with similar complexity.
While our model does take order into order, the temporal distance between events is not considered. Another direction for improvement could be factoring in more complex relations between events, such as multiple dependencies or mutual exclusion. We can also leverage Transformers in more complex systems to better encode each user and what they represent to the system instead of using them directly to predict the next item. 

\section*{Acknowledgment}
We would like to thank Maktabkhooneh, the online learning platform, for their collaboration and providing the dataset to us. 
\bibliographystyle{ieeetr}
\bibliography{references}

\begin{thebibliography}{10}

\bibitem{10.1145/245108.245121}
P.~Resnick and H.~R. Varian, ``Recommender systems,'' {\em Commun. ACM},
  vol.~40, p.~56–58, Mar. 1997.

\bibitem{cheng2016wide}
H.-T. Cheng, L.~Koc, J.~Harmsen, T.~Shaked, T.~Chandra, H.~Aradhye,
  G.~Anderson, G.~Corrado, W.~Chai, M.~Ispir, R.~Anil, Z.~Haque, L.~Hong,
  V.~Jain, X.~Liu, and H.~Shah, ``Wide and deep learning for recommender
  systems,'' in {\em Proceedings of the 1st Workshop on Deep Learning for
  Recommender Systems}, DLRS 2016, (New York, NY, USA), p.~7–10, Association
  for Computing Machinery, 2016.

\bibitem{10.1145/2959100.2959190}
P.~Covington, J.~Adams, and E.~Sargin, ``Deep neural networks for youtube
  recommendations,'' in {\em Proceedings of the 10th ACM Conference on
  Recommender Systems}, RecSys '16, (New York, NY, USA), p.~191–198,
  Association for Computing Machinery, 2016.

\bibitem{10.1145/3159652.3159656}
J.~Tang and K.~Wang, ``Personalized top-n sequential recommendation via
  convolutional sequence embedding,'' in {\em Proceedings of the Eleventh ACM
  International Conference on Web Search and Data Mining}, WSDM '18, (New York,
  NY, USA), p.~565–573, Association for Computing Machinery, 2018.

\bibitem{10.1145/3159652.3159668}
X.~Chen, H.~Xu, Y.~Zhang, J.~Tang, Y.~Cao, Z.~Qin, and H.~Zha, ``Sequential
  recommendation with user memory networks,'' in {\em Proceedings of the
  Eleventh ACM International Conference on Web Search and Data Mining}, WSDM
  '18, (New York, NY, USA), p.~108–116, Association for Computing Machinery,
  2018.

\bibitem{ijcai2019-883}
S.~Wang, L.~Hu, Y.~Wang, L.~Cao, Q.~Z. Sheng, and M.~Orgun, ``Sequential
  recommender systems: Challenges, progress and prospects,'' in {\em
  Proceedings of the Twenty-Eighth International Joint Conference on Artificial
  Intelligence, {IJCAI-19}}, pp.~6332--6338, International Joint Conferences on
  Artificial Intelligence Organization, 7 2019.

\bibitem{kang2018selfattentive}
W.~Kang and J.~J. McAuley, ``Self-attentive sequential recommendation,'' in
  {\em {IEEE} International Conference on Data Mining, {ICDM} 2018, Singapore,
  November 17-20, 2018}, pp.~197--206, {IEEE} Computer Society, 2018.

\bibitem{10.1145/3109859.3109877}
T.~Donkers, B.~Loepp, and J.~Ziegler, ``Sequential user-based recurrent neural
  network recommendations,'' in {\em Proceedings of the Eleventh ACM Conference
  on Recommender Systems}, RecSys '17, (New York, NY, USA), p.~152–160,
  Association for Computing Machinery, 2017.

\bibitem{hidasi2016sessionbased}
B.~Hidasi, A.~Karatzoglou, L.~Baltrunas, and D.~Tikk, ``Session-based
  recommendations with recurrent neural networks,'' in {\em 4th International
  Conference on Learning Representations, {ICLR} 2016, San Juan, Puerto Rico,
  May 2-4, 2016, Conference Track Proceedings} (Y.~Bengio and Y.~LeCun, eds.),
  2016.

\bibitem{10.1145/3269206.3271761}
B.~Hidasi and A.~Karatzoglou, ``Recurrent neural networks with top-k gains for
  session-based recommendations,'' in {\em Proceedings of the 27th ACM
  International Conference on Information and Knowledge Management}, CIKM '18,
  (New York, NY, USA), p.~843–852, Association for Computing Machinery, 2018.

\bibitem{vaswani2017attention}
A.~Vaswani, N.~Shazeer, N.~Parmar, J.~Uszkoreit, L.~Jones, A.~N. Gomez, L.~u.
  Kaiser, and I.~Polosukhin, ``Attention is all you need,'' in {\em Advances in
  Neural Information Processing Systems} (I.~Guyon, U.~V. Luxburg, S.~Bengio,
  H.~Wallach, R.~Fergus, S.~Vishwanathan, and R.~Garnett, eds.), vol.~30,
  Curran Associates, Inc., 2017.

\bibitem{devlin2019bert}
J.~Devlin, M.~Chang, K.~Lee, and K.~Toutanova, ``{BERT:} pre-training of deep
  bidirectional transformers for language understanding,'' in {\em Proceedings
  of the 2019 Conference of the North American Chapter of the Association for
  Computational Linguistics: Human Language Technologies, {NAACL-HLT} 2019,
  Minneapolis, MN, USA, June 2-7, 2019, Volume 1 (Long and Short Papers)}
  (J.~Burstein, C.~Doran, and T.~Solorio, eds.), pp.~4171--4186, Association
  for Computational Linguistics, 2019.

\bibitem{sun2019bert4rec}
F.~Sun, J.~Liu, J.~Wu, C.~Pei, X.~Lin, W.~Ou, and P.~Jiang, ``Bert4rec:
  Sequential recommendation with bidirectional encoder representations from
  transformer,'' in {\em Proceedings of the 28th ACM International Conference
  on Information and Knowledge Management}, CIKM '19, (New York, NY, USA),
  p.~1441–1450, Association for Computing Machinery, 2019.

\bibitem{chen2019behavior}
Q.~Chen, H.~Zhao, W.~Li, P.~Huang, and W.~Ou, ``Behavior sequence transformer
  for e-commerce recommendation in alibaba,'' in {\em Proceedings of the 1st
  International Workshop on Deep Learning Practice for High-Dimensional Sparse
  Data}, DLP-KDD '19, (New York, NY, USA), Association for Computing Machinery,
  2019.

\bibitem{10.1145/3340531.3411954}
K.~Zhou, H.~Wang, W.~X. Zhao, Y.~Zhu, S.~Wang, F.~Zhang, Z.~Wang, and J.-R.
  Wen, ``S3-rec: Self-supervised learning for sequential recommendation with
  mutual information maximization,'' in {\em Proceedings of the 29th ACM
  International Conference on Information and Knowledge Management}, CIKM '20,
  (New York, NY, USA), p.~1893–1902, Association for Computing Machinery,
  2020.

\bibitem{10.1145/3383313.3412258}
L.~Wu, S.~Li, C.-J. Hsieh, and J.~Sharpnack, ``Sse-pt: Sequential
  recommendation via personalized transformer,'' in {\em Fourteenth ACM
  Conference on Recommender Systems}, RecSys '20, (New York, NY, USA),
  p.~328–337, Association for Computing Machinery, 2020.

\bibitem{Schafer2007}
J.~B. Schafer, D.~Frankowski, J.~Herlocker, and S.~Sen, {\em Collaborative
  Filtering Recommender Systems}, pp.~291--324.
\newblock Berlin, Heidelberg: Springer Berlin Heidelberg, 2007.

\bibitem{NIPS2007_d7322ed7}
A.~Mnih and R.~R. Salakhutdinov, ``Probabilistic matrix factorization,'' in
  {\em Advances in Neural Information Processing Systems} (J.~Platt, D.~Koller,
  Y.~Singer, and S.~Roweis, eds.), vol.~20, Curran Associates, Inc., 2008.

\bibitem{10.5555/1005332.1044709}
P.~O. Hoyer, ``Non-negative matrix factorization with sparseness constraints,''
  {\em J. Mach. Learn. Res.}, vol.~5, p.~1457–1469, Dec. 2004.

\bibitem{10.1145/3038912.3052569}
X.~He, L.~Liao, H.~Zhang, L.~Nie, X.~Hu, and T.-S. Chua, ``Neural collaborative
  filtering,'' in {\em Proceedings of the 26th International Conference on
  World Wide Web}, WWW '17, (Republic and Canton of Geneva, CHE), p.~173–182,
  International World Wide Web Conferences Steering Committee, 2017.

\bibitem{10.1145/3097983.3098077}
X.~Li and J.~She, ``Collaborative variational autoencoder for recommender
  systems,'' in {\em Proceedings of the 23rd ACM SIGKDD International
  Conference on Knowledge Discovery and Data Mining}, KDD '17, (New York, NY,
  USA), p.~305–314, Association for Computing Machinery, 2017.

\bibitem{Xiao_Liang_Shen_Meng_2019}
T.~Xiao, S.~Liang, W.~Shen, and Z.~Meng, ``Bayesian deep collaborative matrix
  factorization,'' {\em Proceedings of the AAAI Conference on Artificial
  Intelligence}, vol.~33, pp.~5474--5481, Jul. 2019.

\bibitem{varasteh2021improved}
M.~Varasteh, M.~S. Nejad, H.~Moradi, M.~A. Sadeghi, and A.~Kalhor, ``An
  improved hybrid recommender system: Integrating document context-based and
  behavior-based methods,'' {\em CoRR}, 2021.

\bibitem{10.1145/2959100.2959165}
D.~Kim, C.~Park, J.~Oh, S.~Lee, and H.~Yu, ``Convolutional matrix factorization
  for document context-aware recommendation,'' in {\em Proceedings of the 10th
  ACM Conference on Recommender Systems}, RecSys '16, (New York, NY, USA),
  p.~233–240, Association for Computing Machinery, 2016.

\bibitem{DBLP:conf/icdm/KangFWM17}
W.~Kang, C.~Fang, Z.~Wang, and J.~J. McAuley, ``Visually-aware fashion
  recommendation and design with generative image models,'' in {\em 2017 {IEEE}
  International Conference on Data Mining, {ICDM} 2017, New Orleans, LA, USA,
  November 18-21, 2017} (V.~Raghavan, S.~Aluru, G.~Karypis, L.~Miele, and
  X.~Wu, eds.), pp.~207--216, {IEEE} Computer Society, 2017.

\bibitem{NIPS2013_b3ba8f1b}
A.~van~den Oord, S.~Dieleman, and B.~Schrauwen, ``Deep content-based music
  recommendation,'' in {\em Advances in Neural Information Processing Systems}
  (C.~J.~C. Burges, L.~Bottou, M.~Welling, Z.~Ghahramani, and K.~Q. Weinberger,
  eds.), vol.~26, Curran Associates, Inc., 2013.

\bibitem{10.3389/fams.2019.00044}
M.~Schedl, ``Deep learning in music recommendation systems,'' {\em Frontiers in
  Applied Mathematics and Statistics}, vol.~5, p.~44, 2019.

\bibitem{8734976}
R.~E. Nakhli, H.~Moradi, and M.~A. Sadeghi, ``Movie recommender system based on
  percentage of view,'' in {\em 2019 5th Conference on Knowledge Based
  Engineering and Innovation (KBEI)}, pp.~656--660, 2019.

\bibitem{10.1145/1772690.1772773}
S.~Rendle, C.~Freudenthaler, and L.~Schmidt-Thieme, ``Factorizing personalized
  markov chains for next-basket recommendation,'' in {\em Proceedings of the
  19th International Conference on World Wide Web}, WWW '10, (New York, NY,
  USA), p.~811–820, Association for Computing Machinery, 2010.

\bibitem{9018047}
N.~Wen and F.~Zhang, ``Extended factorization machines for sequential
  recommendation,'' {\em IEEE Access}, vol.~8, pp.~41342--41350, 2020.

\bibitem{10.1145/3240323.3240356}
R.~Pasricha and J.~McAuley, ``Translation-based factorization machines for
  sequential recommendation,'' in {\em Proceedings of the 12th ACM Conference
  on Recommender Systems}, RecSys '18, (New York, NY, USA), p.~63–71,
  Association for Computing Machinery, 2018.

\bibitem{10.1145/3109859.3109900}
T.~X. Tuan and T.~M. Phuong, ``3d convolutional networks for session-based
  recommendation with content features,'' in {\em Proceedings of the Eleventh
  ACM Conference on Recommender Systems}, RecSys '17, (New York, NY, USA),
  p.~138–146, Association for Computing Machinery, 2017.

\bibitem{10.1145/3404835.3462908}
S.~Zhang, D.~Yao, Z.~Zhao, T.-S. Chua, and F.~Wu, ``Causerec: Counterfactual
  user sequence synthesis for sequential recommendation,'' in {\em Proceedings
  of the 44th International ACM SIGIR Conference on Research and Development in
  Information Retrieval}, SIGIR '21, (New York, NY, USA), p.~367–377,
  Association for Computing Machinery, 2021.

\end{thebibliography}
\end{document}